\documentclass[a4paper,11pt]{article}
\pdfoutput=1 
\usepackage{subfigure}
\usepackage{jcappub}
\usepackage[T1]{fontenc}
\title{Cosmography of Cardassian model}
\author{Yu.L. Bolotin,\note{Corresponding author.}}
\author{M.I.Konchatnyi,}
\author{O.A. Lemets}
\author{and L.G. Zazunov}
\affiliation{A.I.Akhiezer Institute for Theoretical Physics, National Science Center ''Kharkov Institute of Physics and Technology'',\\Akademicheskaya Str. 1, 61108 Kharkov, Ukraine}
\emailAdd{ybolotin@gmail.com}
\abstract{The parameters of any model that satisfies the cosmological principle (the universe is homogeneous and isotropic on large scale), can be expressed through cosmographic  parameters. In this paper, we perform this procedure for the Cardassian model. We demonstrate a number of advantages of the approach used before traditional methods.}
\arxivnumber{}
\begin{document}
\maketitle
\flushbottom
\section{Introduction}
According to the current cosmological paradigm, we live in a flat, rapidly expanding Universe \cite{1,2,3,4}. However, the physical origin of cosmic acceleration is still the greatest mystery to science \cite{5}. The Universe filled with ''ordinary'' components (matter and radiation) should eventually slow down its expansion. The search for a solution to this problem is conducted in two directions, based on the following assumptions:
\begin{enumerate}
  \item Up to 75\% of the energy density of the Universe exists in the form of an unknown substance (commonly called dark energy) with a large negative pressure which ensures the accelerated expansion.
  \item The general relativity theory (GR) needs to be revised on cosmological scales.
\end{enumerate}

The standard cosmological model (SCM) \cite{6} represents an extremely successful implementation of the first direction. Effectiveness of the SCM was so high that for its adequate description even the special term ''cosmic concordance'' was introduced \cite{7}. A fairly simple model with a small number of parameters makes it possible to describe giant arrays of observational data at the level of 5\% \cite{8,9,10}.

However, unlike fundamental theories, physical models reflect only our current understanding of a given process or phenomenon for which they were created. Gradually growing internal problems and inevitable contradictions with observations \cite{11,12,13,14} stimulate the search for more and more new models. In particular, it is tempting to explain the accelerated expansion of the universe without attracting exotic dark components. Of course, this can be done only by modifying the equations of general relativity. In other words, any ''innovation'' activity in cosmology can be interpreted in terms of modifying either the left or the right side of Einstein's equations. SCM achieves agreement with cosmological observations by including dark energy and dark matter in the energy-momentum tensor. The focus will be on the so-called Cardassian model (CM) \cite{15,16}, which allows us to describe the accelerated expansion of a spatially flat universe without involving dark energy. Accelerated expansion is achieved by modifying the Friedmann equation $H^{2} =\rho \, \to H^{2} =g\left(\rho \right)$, where the density $\rho $ includes only ''ordinary'' components: matter and radiation. The function $g\left(\rho \right)$ is chosen in such a way that in the early Universe (for $z\gg 1$) it reproduces the standard Friedmann equation, while for $z\gg 1$ it generates the accelerated expansion of the Universe. In the simplest version, the CM is characterized by only two parameters. Our goal is to construct a cosmography of the CM.

The cosmography \cite{17,18,19,20,20a,21} represents an approach entirely based on the cosmological principle, stating that the Universe is homogeneous and isotropic on scales larger than a hundred megaparsecs. It allows us to select from whole possible variety of models describing the Universe a narrow set of homogeneous and isotropic models. The cosmological principle enables us to build the metrics and make first steps towards interpretation of the cosmological observations. The cosmography is just the kinematics of cosmological expansion. In order to build the key characteristic -- time dependence of the scale factor $a(t)$  -- one needs to take equations of motion (the Einstein's equations) and make an assumption about material content of the Universe, which allows constructing the energy-momentum tensor. The cosmography is efficient because it allows testing any cosmological model which does not contradict the cosmological principle. Modifications of General Relativity or introduction of new components (such as DM and DE) evidently change the dependence $a(t)$ but do not affect relations between the kinematic characteristics.

To accomplish this goal, one must:
\begin{enumerate}
  \item Express the model parameters through cosmographic parameters.
  \item Find admissible (consistent with observational data) intervals of variation of these parameters.
  \item Analyze the relationship to the results obtained within other cosmological models.
\end{enumerate}

We emphasize that all our results presented below are exact, being derived from identical transformations.

\section{Cardassian model}
As an alternative explanation for the observed accelerated expansion of the universe Freese and Lewis \cite{15} proposed a modified version of the first Friedmann equation
\begin{equation} \label{card-model__1_}
H^{2} =g\left(\rho _{m} \right),
\end{equation}
where the energy density  $\rho _{m} $  of a flat universe includes only nonrelativistic matter (both baryon and dark) and radiation, but does not contain dark energy. In what follows, we restrict to the simplest version of the CM, which uses the additional power law on the right-hand side of the Friedmann equation
\begin{equation} \label{card-model__2_}
H^{2} =A\rho _{m} +B\rho _{m}^{n}.
\end{equation}
In the standard FLRW cosmology coefficient $B=0$. Therefore, we must choose $A=\frac{8\pi G}{3} $.

Suppose that the universe is filled only with non-relativistic matter. In this case, with the dominance of the second term (high densities, late Universe): $H\propto \rho _{m}^{n/2} \propto a^{-3n/2} ,\, \dot{a}\propto a^{-3n/2+1}, \, a\propto t^{2/3n} $. Consequently, accelerated expansion can be realized  for $n<2/3$. At the upper boundary for $n=2/3$ we have $a\propto t$ and $\ddot{a}=0$; for $n=1/3$ we have $a\propto t^{2} $ (the acceleration is constant). For $n>1/3$  the acceleration is diminishing in time, while for $n<1/3$ the acceleration is increasing. It is interesting to note that if $n=2/3$ we have $H^{2} \propto a^{-2} $: in a flat Universe term similar to a curvature term is generated  by matter.

Let's  represent the energy density of the CM in the form of a sum of densities of ordinary matter $\rho _{m} $ and components with a density $\rho _{x} =\rho _{m}^{n} $ such that $H^{2} \propto \rho _{m} +\rho _{x}$. As we saw above, in the case of dominance of the additional term in the Friedmann equation one has $a\propto t^{2/3n}$. Since $a\propto t^{\frac{2}{3\left(w+1\right)} } $ we find that the parameter of the equation of state $p_{x} =w_{x} \rho _{x} $ is
\begin{equation} \label{card-model__3_}
w_{x} =n-1.
\end{equation}
This relation holds for an arbitrary one-component fluid with $w_{x} =const$. In this case
\begin{equation} \label{card-model__4_}
\frac{d\rho _{x} }{dz} =3\rho _{x} \frac{1+w_{x} (z)}{1+z}.
\end{equation}
Taking into account that $\rho _{x} =\rho _{m}^{n} =(1+z)^{3\left(w_{x} +1\right)n} $ and substituting this into \eqref{card-model__4_}, we obtain
\begin{equation} \label{card-model__5_}
w_{x} =n-1.
\end{equation}
For $0<n<2/3$
\begin{equation} \label{card-model__6_}
-1<w_{x} <-1/3.
\end{equation}
As you would expect, this values range of parameter $ w_ {x} $ generates a negative pressure, responsible for the late time accelerated expansion of the universe. As one would expect, this interval of parameter values generates a negative pressure, responsible for the accelerated expansion of the universe at late times.

\section{Cosmography - short review}
In order to make more detailed description of kinematics of cosmological expansion it is useful to consider the extended set of the parameters which includes the Hubble parameter $H(t)\equiv \frac{\dot{a}}{a} $, and higher order time derivatives of the scale factor \cite{17,18}  deceleration parameter $q\equiv -C_{2} $, jerk parameter $j(t)\equiv C_{3} $ snap parameter $s(t)\equiv C_{4} $  etc, where
\begin{equation} \label{card-model__7_}
C_{n} \equiv \frac{1}{a} \frac{d^{n} a}{dt^{n} } H^{-n}.
\end{equation}
Note that all cosmological parameters, except for the deceleration parameter, are dimensionless.

We give a number of useful relations for the deceleration parameter \cite{21}
\begin{equation} \label{card-model__8_}
\begin{array}{l}
{q(t)=\frac{d}{dt} \left(\frac{1}{H} \right)-1;} \\
{q(z)=\frac{1+z}{H} \frac{dH}{dz} -1;} \\
{q(z)=\frac{1}{2} (1+z)\frac{1}{H^{2} } \frac{dH^{2} }{dz} -1;} \\
{q(z)=\frac{1}{2} \frac{d\ln H^{2} }{d\ln (1+z)} -1;} \\
{q(z)=\frac{d\ln H}{dz} (1+z)-1.} \\
\end{array}
\end{equation}
Derivatives $\frac{d^{n} H}{dz^{n} } $  can be expressed through the deceleration parameter and other cosmographic parameters as follows:
\begin{equation} \label{card-model__9_}
\begin{array}{l}
{\frac{dH}{dz} =\frac{1+q}{1+z} H;} \\
{\frac{d^{2} H}{dz^{2} } =\frac{j-q^{2} }{\left(1+z\right)^{2} } H;} \\
{\frac{d^{3} H}{dz^{3} } =\frac{H}{(1+z)^{3} } \left(3q^{2} +3q^{3} -4qj-3j-s\right);} \\
{\frac{d^{4} H}{dz^{4} } =\frac{H}{(1+z)^{4} } \left(-12q^{2} -24q^{3} -15q^{4} +32qj+25q^{2} j+7qs+12j-4j^{2} +8s+l\right).}
\end{array}
\end{equation}

Derivatives $\frac{d^{(i)} H^{2} }{dz^{(i)} } ,\; i=1,2,3,4$  in terms of cosmographic  parameters have the form
\begin{equation} \label{card-model__11_}
\begin{array}{l}
{\frac{d(H^{2} )}{dz} =\frac{2H^{2} }{1+z} (1+q);} \\
{\frac{d^{2} (H^{2} )}{dz^{2} } =\frac{2H^{2} }{(1+z)^{2} } (1+2q+j);} \\
{\frac{d^{3} (H^{2} )}{dz^{3} } =\frac{2H^{2} }{(1+z)^{3} } (-qj-s);} \\
{\frac{d^{4} (H^{2} )}{dz^{4} } =\frac{2H^{2} }{(1+z)^{4} } (4qj+3qs+3q^{2} j-j^{2} +4s+l).}
\end{array}
\end{equation}
The current values of deceleration and jerk parameters in terms of $N=-\ln (1+z)$ are
\begin{equation} \label{card-model__12_}
\begin{array}{l}
{q_{0} =\left. -\frac{1}{H^{2} } \left\{\frac{1}{2} \frac{d\left(H^{2} \right)}{dN} +H^{2} \right\}\right|_{N=0} ,\quad } \\
{j_{0} =\left. \frac{1}{2H^{2} } \frac{d^{2} \left(H^{2} \right)}{dN^{2} } +\frac{3}{2H^{2} } \frac{d\left(H^{2} \right)}{dN} +1\right|_{N=0} .}
\end{array}
\end{equation}
The derivatives of the Hubble parameter with respect to time can also be expressed in terms of cosmographic  parameters
\begin{equation} \label{card-model__13_}
\begin{array}{l}
{\dot{H}=-H^{2} (1+q);} \\
{\ddot{H}=H^{3} \left(j+3q+2\right);} \\
{\stackrel{...}{H}=H^{4} \left[s-4j-3q(q+4)-6\right];} \\
{\stackrel{....}{H}=H^{5} \left[l-5s+10\left(q+2\right)j+30(q+2)q+24\right].}
\end{array}
\end{equation}
Let $C_{n} \equiv \gamma _{n} \frac{a^{(n)} }{aH^{n} } $ where $a^{(n)} $ is the -n-th derivative of the scale factor with respect to time, $n\ge 2$ and $\gamma _{2} =-1, \; \gamma _{n} =1$ for $n>2$.  Then for the derivatives of the parameters $C_{n} $ with respect to the redshift, we have relations
\begin{equation} \label{card-model__14_}
(1+z)\frac{dC_{n} }{dz} =-\frac{\gamma _{n} }{\gamma _{n+1} } C_{n+1} +C_{n} -nC_{n} (1+q)
\end{equation}
Using this relationship, one can express the higher cosmographic parameters through the lower ones and their derivatives:
\begin{equation} \label{card-model__15_}
\begin{array}{l}
{j=-q+(1+z)\frac{dq}{dz} +2q(1+q),} \\
{s=j-3j(1+q)-(1+z)\frac{dj}{dz} ,} \\ {l=s-4s(1+q)-(1+z)\frac{ds}{dz} ,} \\
{m=l-5l(1+q)-(1+z)\frac{dl}{dz} .}
\end{array}
\end{equation}
We then proceed with computing the deceleration parameter in the CM \cite{22,23}. For this, one naturally has to go beyond the limits of cosmography and turn to the dynamics described by the Friedmann equations. There is a complete analogy with classical mechanics. Kinematics describes motion ignoring the forces that generate this motion. To calculate the acceleration, Newton's laws are necessary, i.e. dynamics. Using
\begin{equation} \label{card-model__16_}
\begin{array}{l} {q(z)=\frac{1}{2} (1+z)\frac{1}{H^{2} } \frac{dH^{2} }{dz} -1=\frac{1}{2} \frac{(1+z)}{E^{2} (z)} \frac{dE^{2} (z)}{dz} -1,} \\ {E^{2} \equiv \frac{H^{2} }{H_{0}^{2} } =\Omega _{m0} (1+z)^{3} +\left(1-\Omega _{m0} \right)(1+z)^{3n} } \end{array}
\end{equation}
one finds
\begin{equation} \label{card-model__17_}
\begin{array}{l}
{q(z)=\frac{1}{2} -\frac{3}{2} \left(1-n\right)\frac{\kappa (z)}{\left(1+\kappa (z)\right)} ,\quad } \\
{\kappa (z)\equiv \left(\frac{1-\Omega _{m0} }{\Omega _{m0} } \right)\left(1+z\right)^{-3\left(1-n\right)} } \end{array}
\end{equation}
For the current value of the deceleration parameter $q_{0} =q\left(z=0\right)$, we get
\begin{equation} \label{card-model__18_}
q_{0} =\frac{1}{2} -\frac{3}{2} (1-n)\left(1-\Omega _{m0} \right)
\end{equation}
For $n=0$ and $\Omega _{m0} =0$ the relation \eqref{card-model__18_} correctly reproduces the value of the deceleration parameter $q=-1$ for the cosmological constant.

\section{Cosmography of cardassian model}
Dunajski and Gibbons \cite{24} proposed an original approach for testing cosmological models which satisfy the cosmological principle. Implementation of the method relies on the following sequence of steps:
\begin{enumerate}
  \item The first Friedmann equation is transformed to the ODE for the scale factor. To achieve this, the conservation equation for each component included in the model is used to find the dependence of the energy density on the scale factor.
  \item The resulting equation is differentiated (with respect to cosmological time) as many times as the number of free parameters of the model.
  \item Temporal derivatives of the scale factor are expressed through cosmographic parameters.
  \item All free parameters of the model are expressed through cosmological parameters by solving the resulting system of linear algebraic equations
\end{enumerate}

The above procedure can be made more universal and effective when choosing the system of Friedmann equations for the Hubble parameter $H$ and its time derivative $\dot{H}$ as a starting point.  By differentiating the equation for $\dot{H}$ the required number of times (this number is determined by the number of free parameters of the model), we obtain a system of equations that includes higher time derivatives of the Hubble parameter $\ddot{H},\dddot{H},\ddddot{H}...$ These derivatives are directly related to the cosmological parameters by the relations \eqref{card-model__9_}. We implement this procedure for the CM.

\noindent The evolution of CM is described by the system of equations
\begin{equation} \label{card-model__19_}
H^{2} =A\rho _{m} +B\rho _{m}^{n}
\end{equation}
\begin{equation} \label{card-model__20_}
\dot{\rho }_{m} +3H\rho _{m} =0
\end{equation}
Differentiating Eq. \eqref{card-model__19_} with respect to the cosmological time and using \eqref{card-model__20_} we transform the system \eqref{card-model__19_}-\eqref{card-model__20_} to the form
\begin{equation} \label{card-model__21_}
H^{2} =A\rho _{m} +B\rho _{m}^{n}
\end{equation}
\begin{equation} \label{card-model__22_}
-\frac{2}{3} \dot{H}=A\rho _{m} +Bn\rho _{m}^{n}
\end{equation}
The solutions  of  this system are given by
\begin{equation} \label{card-model__23_}
\rho _{m} =-\frac{nH^{2} +\frac{2}{3} \dot{H}}{A(1-n)}
\end{equation}
\begin{equation} \label{card-model__24_}
B=\frac{H^{2} +\frac{2}{3} \dot{H}}{\rho ^{n} (1-n)}
\end{equation}
To calculate the parameter $n$, we need an expression for $\ddot{H}$,
\begin{equation} \label{card-model__25_}
\frac{2}{9} \frac{\ddot{H}}{H} =A\rho _{m} +n^{2} B\rho _{m}^{n}
\end{equation}
Substituting the above solutions \eqref{card-model__23_} and \eqref{card-model__24_} for $\rho $ and $B$  into this expression, we obtain
\begin{equation} \label{card-model__26_}
\frac{2}{9} \frac{\ddot{H}}{H^{3} } =-n+\frac{2}{3} \frac{\dot{H}}{H^{2} } (1+n)
\end{equation}
Hence for the parameter  $n$ we find
\begin{equation} \label{card-model__27_}
n=-\frac{\frac{2}{3} \left(\frac{1}{3} \frac{\ddot{H}}{H^{3} } +\frac{\dot{H}}{H^{2} } \right)}{1+\frac{2}{3} \frac{\dot{H}}{H^{2} } }
\end{equation}
Expressions \eqref{card-model__24_} and \eqref{card-model__27_} allow us to express  the parameters of CM through cosmographic parameters. Using known expressions for time derivatives of the Hubble parameter in terms of cosmological parameters \eqref{card-model__13_}, we obtain
\begin{equation} \label{card-model__28_}
\frac{B\rho _{m}^{n} }{H^{2} } =\frac{1}{3}\frac{\left(1-2q\right)}{1-n}
\end{equation}
\begin{equation} \label{card-model__29_}
n=\frac{2}{3} \frac{j-1}{2q-1}
\end{equation}
Note that the parameters $B$ and $n$ are constants, which was explicitly used in deriving the above relations. We verify that the solutions \eqref{card-model__28_} and \eqref{card-model__29_} agree with this condition. The requirement $\dot{B}=0$ is transformed into \eqref{card-model__26_} and, consequently, it is consistent with the above expression for $n$. The constancy of the parameters allows us to compute them for the values of the cosmological parameters at any time. Since the main body of information about cosmological parameters refers to the current time $t_{0} $, the relations \eqref{card-model__28_}, \eqref{card-model__29_} can be put in the form
\begin{equation} \label{card-model__30_}
\frac{B\rho _{0}^{n} }{H_{0}^{2} } =\frac{1}{3} \frac{\left(1-2q_{0} \right)}{1-n}
\end{equation}
\begin{equation} \label{card-model__31_}
n=\frac{2}{3} \frac{j_{0} -1}{2q_{0} -1}
\end{equation}
Otherwise, we must treat the time-dependent solution \eqref{card-model__23_} for the density $\rho _{m} $.  It can be represented in the form
\begin{equation} \label{card-model__32_}
\frac{\rho }{\rho _{c} } =\frac{-n+\frac{2}{3} \left(1+q\right)}{1-n} ,\quad \rho _{c} \equiv \frac{3H^{2} }{8\pi G}
\end{equation}
The current density in CM can be found by substitution $q\to q_{0} ,\; H\to H_{0} $.

It is interesting to note that the expression \eqref{card-model__29_}  for the parameter $n$   coincides exactly with the parameter $s$, one of the so-called statefinder parameters $\left\{r,s\right\}$  \cite{25,26},
\begin{equation} \label{card-model__33_}
r\equiv \frac{\dot{a}}{aH^{^{3} } } ,\quad s=\frac{2}{3} \frac{r-1}{2q-1}
\end{equation}
The coincidence is obvious, since $r\equiv j$. The reason for the coincidence can be explained as follows. In any model with the scale factor $a\propto t^{\alpha } $, there are the simple relations for the cosmographic parameters $q$ and $j$
\begin{equation} \label{card-model__34_}
2q-1=\frac{2-3\alpha }{\alpha } ,\quad j-1=\frac{2-3\alpha }{\alpha ^{2} }
\end{equation}
In CM $a\propto t^{\frac{2}{3n} } $, from which it follows that $s=n$.

Using the expression for $\stackrel{...}{H}$ \eqref{card-model__13_} and the solutions found \eqref{card-model__23_}, \eqref{card-model__24_}, \eqref{card-model__29_}, we obtain an equation relating cosmological parameters
\begin{equation} \label{card-model__35_}
\begin{array}{l}
{s+qj+\left(3n+2\right)j-2q(3n-1)=0,} \\
{n=\frac{2}{3} \frac{j-1}{2q-1} .}
\end{array}
\end{equation}
This fourth order ODE for the scale factor  is equivalent to the Friedmann equation. For $n=0$  Eq. \eqref{card-model__35_} reproduces the known relation \cite{24} between the cosmographic parameters in the LCDM,
\begin{equation} \label{card-model__36_}
s+2(q+j)+qj=0
\end{equation}
We now turn to the dimensionless form of the evolution equation for the scale factor \eqref{card-model__19_}. Coefficients of this equation must be expressed in terms of cosmological parameters. For this using $\rho _{m} =\frac{\rho _{0} }{a^{3} } $ we represent Friedmann equation in the form
\begin{equation} \label{card-model__37_}
\dot{a}^{2} =A\rho _{0} a^{-1} +B\rho _{0}^{n} a^{-3n+2}.
\end{equation}
Passing to the dimensionless time $\tau =H_{0} t$ and substituting in \eqref{card-model__37_},
\begin{equation} \label{card-model__38_}
\rho _{0} =-\frac{nH_{0}^{2} +\frac{2}{3} \dot{H}_{0} }{A(1-n)} ,\quad B=\frac{H_{0}^{2} +\frac{2}{3} \dot{H}_{0} }{\rho _{0}^{n} (1-n)}
\end{equation}
transform  Friedmann equation \eqref{card-model__37_} to the form
\begin{equation} \label{card-model__39_}
\begin{array}{l}
{\frac{da}{d\tau } =\left[F\left(q_{0} ,j_{0} \right)a^{-1} +\Phi \left(q_{0} ,j_{0} \right)a^{-3n+2} \right]^{1/2} ,} \\
{F=-\frac{n-\frac{2}{3} \left(1+q_{0} \right)}{1-n} ,\quad \Phi =\frac{1-\frac{2}{3} \left(1+q_{0} \right)}{1-n} ,} \\
{n=\frac{2}{3} \frac{j_{0} -1}{2q_{0} -1} .}
\end{array}
\end{equation}
This equation should be solved with constraints $n<2/3$ (the condition ensuring the accelerated expansion of the universe) and $n<\frac{2}{3} \left(1+q_{0} \right)$ (the condition ensuring the positivity of the energy density). It is easy to see that the two conditions are consistent, since in the case of the accelerated expansion one has $q<0$.


\begin{figure}[h!]
\centering 
\includegraphics[width=.49\textwidth]{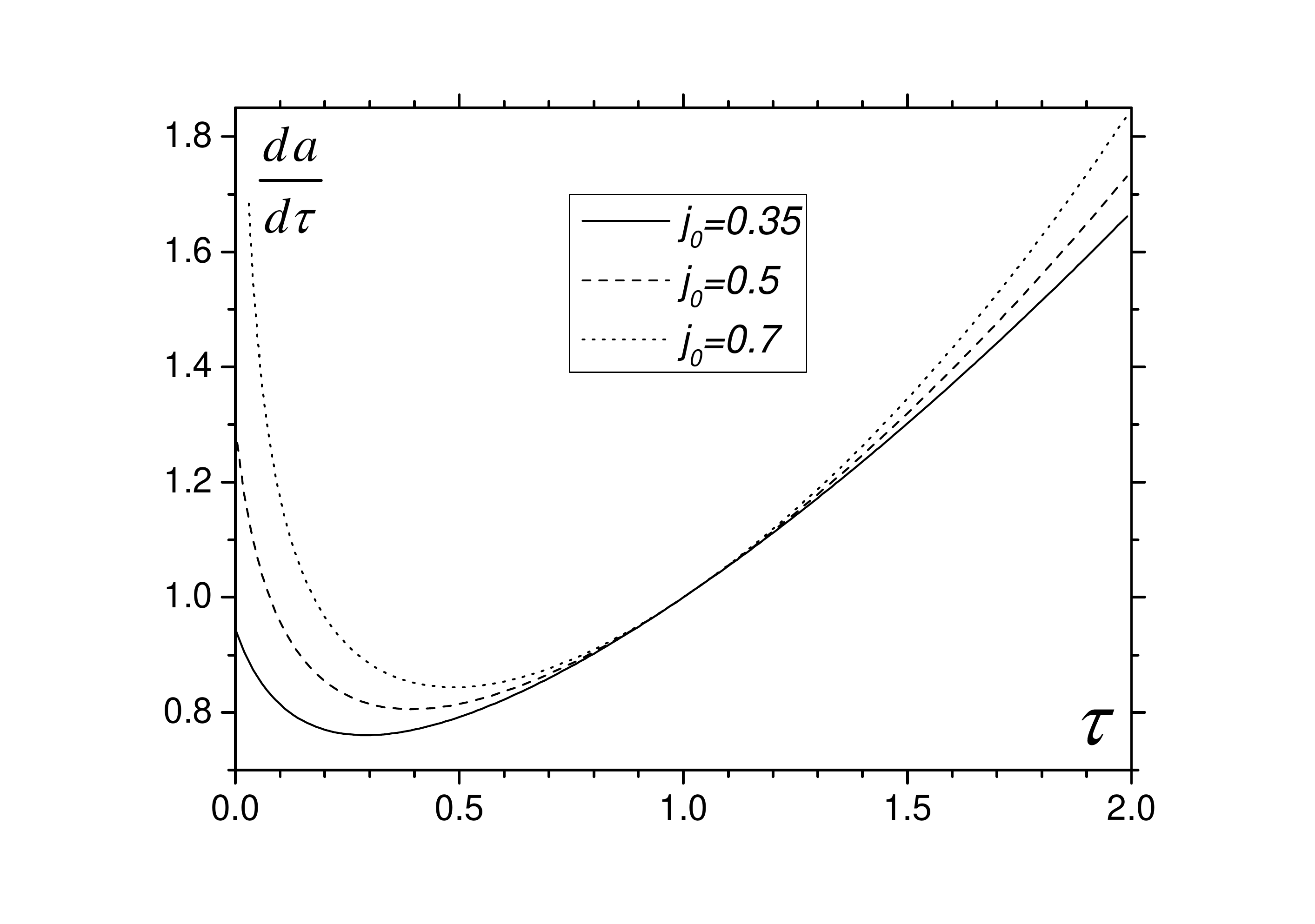}
\hfill
\includegraphics[width=.49\textwidth]{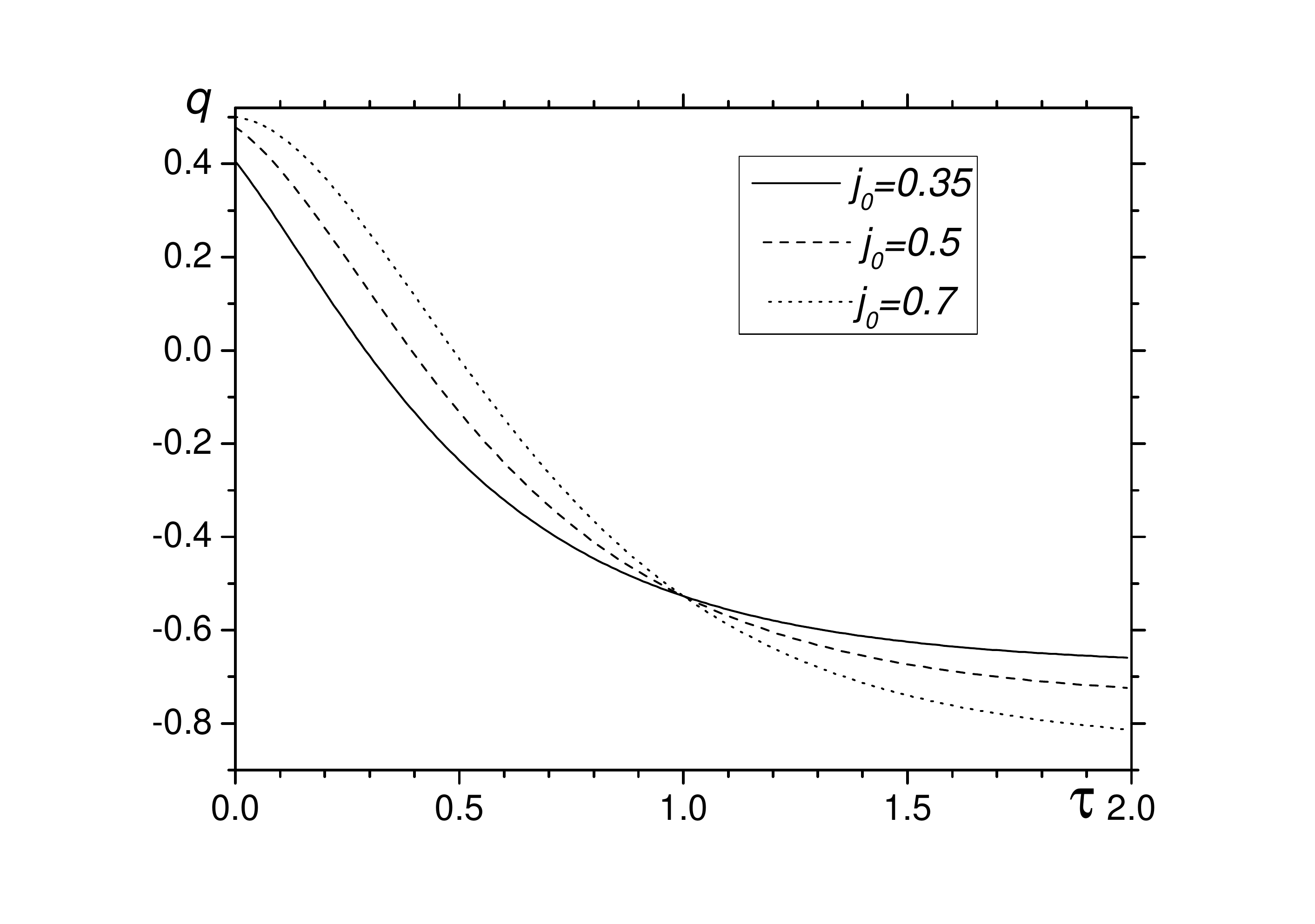}
\caption{\label{fig:1} Demonstration of the transition from delayed to accelerated expansion for different values of the parameter $j_{0} $, with $q_{0} =-0.527$ for all three variants. Left figure: derivative $da/d \tau $  as a function of dimensionless time. Right figure: the deceleration parameter $q$ as a function of dimensionless time.}
\end{figure}

Fig. \ref{fig:1} illustrates the transition from delayed expansion to accelerated for late-time Universe in CM. The growth of the parameter $j_{0} $  leads to a decrease of the parameter $n:$ $j=0.35$,\, $n\approx 0.21$;\,$j=0.5, \, n\approx 016$,\,$j=0.7,\, n\approx 0.097$. If $j\to 1$ (LDCM), as expected, $n\to 0$.

\section{Advantages of cosmographic description}
The considered here method of finding the parameters of cosmological models has some advantages. Let us briefly dwell on them.
\begin{enumerate}
  \item Universality: the method is applicable to any braiding model that satisfies the cosmological principle. This procedure can be generalized to the case of models with interactions between their components \cite{27}.
  \item Reliability: all the derived expressions are exact, as they follow from identical transformations.
  \item The simplicity of the procedure.
  \item Parameters of various models are expressed through a universal set of cosmological parameters. There is no need to introduce additional parameters. Let us illustrate this statement on the following example.

The authors of the CM suggested the following procedure for estimation of the   parameter $B$  \cite{15}. The original CM is described by a set of parameters $\left\{B,n\right\}$. We pass to a new set of parameters $\left\{B,n\right\}\to \left\{z_{eq} ,n\right\}$, where $z_{eq} $ there is a redshift, in which the contributions from the members $A\rho _{m} $ and $B\rho _{m}^{n} $ are compared,
\begin{equation} \label{GrindEQ__40_}
A\rho \left(z_{eq} \right)=B\rho ^{n} \left(z_{eq} \right).
\end{equation}
Since $\rho =\rho _{0} /a^{3} =\rho _{0} (1+z)^{3} $, then
\begin{equation} \label{card-model__41_}
\frac{B}{A} =\rho _{0}^{1-n} \left(1+z_{eq} \right)^{3\left(1-n\right)}.
\end{equation}
Using for the parameter $A$,
\begin{equation} \label{card-model__42_}
A=\frac{H_{0}^{2} }{\rho _{0} } -B\rho _{0}^{n-1},
\end{equation}
obtain
\begin{equation} \label{card-model__43_}
\frac{B\rho _{0}^{n} }{H_{0}^{2} } =\frac{\left(1+z_{eq} \right)^{3\left(1-n\right)} }{1+\left(1+z_{eq} \right)^{3\left(1-n\right)} }.
\end{equation}
The CMB and supernovae data allow to limit the interval of change $z_{eq} $,  $0.3<z_{eq} <1$. Comparing \eqref{card-model__30_} and \eqref{card-model__43_}, we are convinced of the obvious advantage of the cosmographic approach: to find the parameter $B$, we did not have to introduce additional parameters. The dimensionless parameter $\frac{B\rho _{0}^{n} }{H_{0}^{2} } $ is determined by the current value of the fundamental cosmological parameter, the deceleration parameter $q_{0} $. From equating \eqref{card-model__30_} to \eqref{card-model__43_}, we find a function $z_{eq} (n,q_{0} )$ that allows us to estimate the interval of variation of the parameter $n$  corresponding to the interval $0.3<z_{eq} <1$. We see (see Fig. \ref{fig:2}) that this interval includes parameters $n\le 0.2$.

\begin{figure}[h!]
\centering 
\includegraphics[width=.9\textwidth]{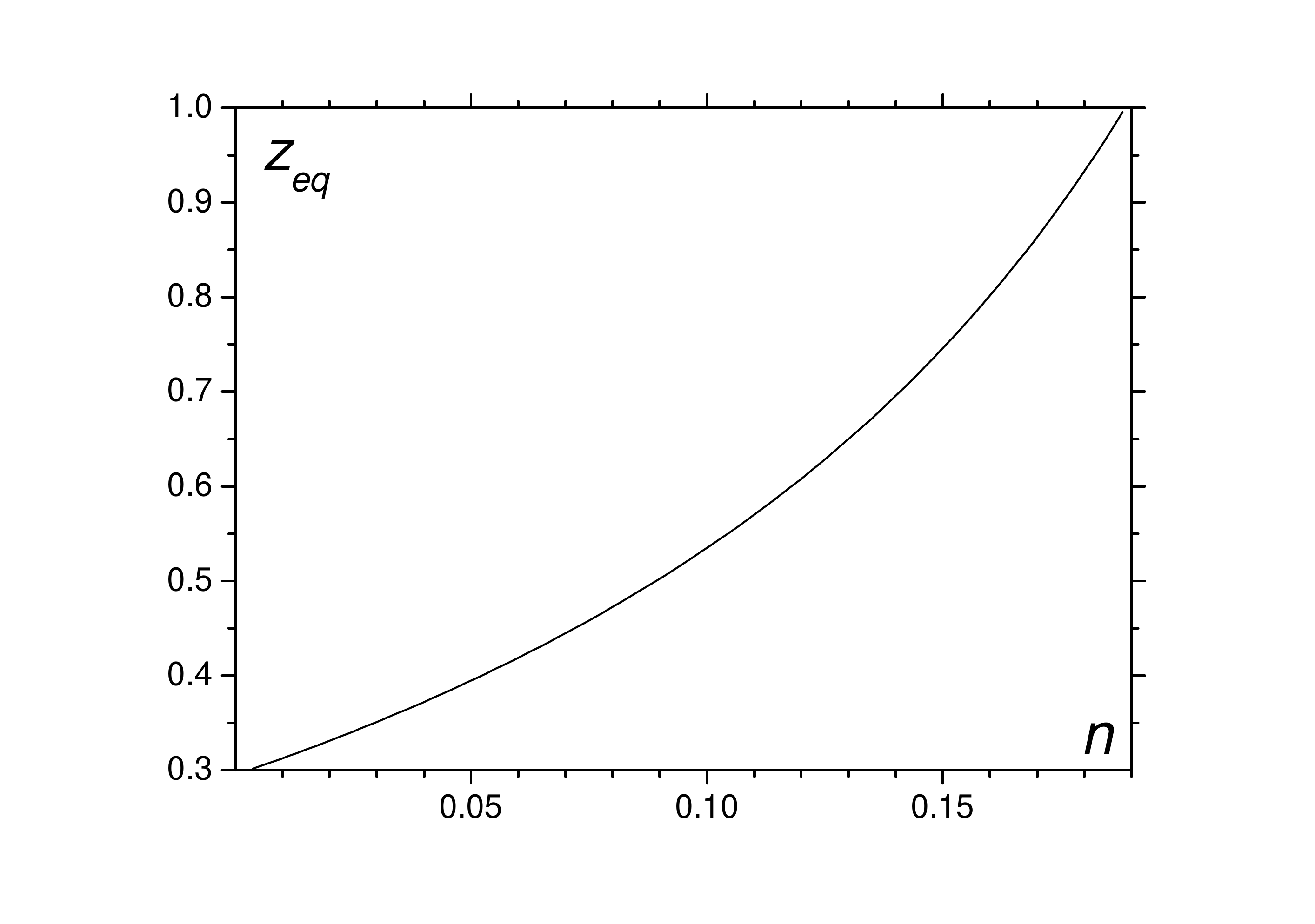}
\caption{\label{fig:2} Function $z_{eq} (n,q_{0} )$ at the value of $q_{0} =-0.527$ for the current deceleration parameter .}
\end{figure}

  \item The method provides an interesting possibility of calculating the highest cosmological parameters from the values of lower parameters known with a better accuracy. For example, Eq. \eqref{card-model__31_} can be used to estimate the parameter $s_{0} $ for known values of $q_{0} $ and $j_{0} $,
\begin{equation} \label{card-model__44_}
\begin{array}{l}
{s_{0} =-q_{0} j_{0} -\left(3n+2\right)j_{0} +2q_{0} (3n-1),} \\
{n=\frac{2}{3} \frac{j_{0} -1}{2q_{0} -1} .}
\end{array}
\end{equation}
In particular, in LCDM $j=1$ and the relation \eqref{card-model__36_} is transformed into $s=-3q-2$. It is easy to see that the cosmographic parameters of the LCDM
\begin{equation} \label{card-model__45_}
\begin{array}{l}
{q=-1+\frac{3}{2} \Omega _{m} ,} \\
{s=1-\frac{9}{2} \Omega _{m} ,}
\end{array}
\end{equation}
exactly satisfy this relationship.
  \item The method presents a simple test for analyzing the compatibility of different models. The analysis consists of two steps. In the first step, the model parameters are expressed through cosmological parameters. The second step consists in finding the intervals of cosmological parameter changes that can be realized within the framework of the considered model. Since the cosmological parameters are universal, only in the case of a nonzero intersection of the obtained intervals, the models are compatible.
\end{enumerate}

\end{document}